\documentclass[12pt]{iopart}
\usepackage{iopams}
\usepackage{bbold}
\usepackage{graphicx}
\usepackage{float}

\begin{document}

\title[]{Quantum Correlations in a Few-Atom Spin-1 Bose-Hubbard Model}

\author[]{B. \c{C}akmak, G. Karpat and Z. Gedik}
\address{Faculty of Engineering and Natural Sciences, Sabanci University, Tuzla, Istanbul 34956, Turkey}
\ead{cakmakb@sabanciuniv.edu}

\begin{abstract}
We study the thermal quantum correlations and entanglement in spin-1 Bose-Hubbard model with two and three particles. While we use negativity to calculate entanglement, more general non-classical correlations are quantified using a new measure based on a necessary and sufficient condition for zero-discord state. We demonstrate that the energy level crossings in the ground state of the system are signalled by both the behavior of thermal quantum correlations and entanglement.
\end{abstract}

\section{Introduction}
Entanglement, being considered as the resource of quantum information science, has been utilized to investigate various properties of condensed matter systems [1,2]. However, it has been discovered that entanglement is not the only kind of useful non-classical correlation present in quantum systems. Numerous quantifiers of quantum correlations have been proposed to reveal the non-classical correlations that cannot be captured by entanglement measures [3].

Quantum phase transitions (QPTs) are critical changes in the ground states of many-body systems when one or more of its physical parameters are continuously changed at absolute zero temperature. These qualitative changes can be signalled by energy level crossings in the ground state of the system. Signatures of QPTs are also present at sufficiently low but experimentally accessible temperatures, where thermal fluctuations are still negligible as compared to quantum fluctuations. In recent years, the techniques of quantum information theory have been implemented for the investigation of quantum critical systems. For instance, several correlation measures have been shown to identify the critical points of QPTs with success in quantum critical spin chains [3-5].

In this work, we analyze quantum correlations in a spin-1 Bose-Hubbard model with two and three particles considering periodic boundary conditions. As a measure of quantum correlations, we use a recently introduced measure for an arbitrary bipartite system based on a necessary and sufficient condition for a zero-discord state in the coherence-vector representation of density matrices [6]. On the other hand, we adopt negativity to measure the amount of entanglement in a quantum state. We demonstrate that both quantum correlations that are more general than entanglement and negativity mark the critical points corresponding to energy level crossings in the ground state of the system. Although we only consider few particle systems in our study, this interesting behavior might have consequences even for actual quantum critical systems, where the number of particles is very large and the energy level crossings really lead to quantum QPTs.

\section{Correlation Measures}
In this section, we briefly review the measures of quantum correlations used in our work. We first introduce a measure of non-classical correlations recently proposed by Zhou et al. based on a necessary and sufficient condition for a zero-discord state [6]. A general bipartite state $\rho^{ab}$ can be expressed in coherence-vector representation as
\begin{eqnarray}
\rho^{ab} &= \frac{1}{mn} I^{a} \otimes I^{b} + \sum_{i=1}^{m^2-1}x_{i}X_{i} \otimes \frac{I^{b}}{2n}
+\frac{I^{a}}{2m} \otimes \sum_{j=1}^{n^2-1}y_{j}Y_{j}  \nonumber \\
     & + \frac{1}{4} \sum_{i=1}^{m^2-1}\sum_{j=1}^{n^2-1} t_{ij} X_{i} \otimes Y_{j},
\end{eqnarray}
where the matrices $\{X_{i}: i=0,1,\cdots,m^{2}-1\}$ and $\{Y_{j}: j=0,1,\cdots,n^{2}-1\}$, satisfying $\textmd{tr}(X_{k}X_{l})=\textmd{tr}(Y_{k}Y_{l})=2\delta_{kl}$, define an orthonormal Hermitian operator basis associated to the subsystems $a$ and $b$, respectively. Here, $I$ is the identity matrix for the specified subsystem. The components of the local Bloch vectors $\vec{x}=\{x_{i}\}$, $\vec{y}=\{y_{j}\}$ and the correlation matrix $T=t_{ij}$ can be obtained as
\begin{eqnarray}
x_{i}  &= \textmd{tr}\rho^{ab}(X_{i} \otimes I^{b})\nonumber, \\
y_{j}  &= \textmd{tr}\rho^{ab}(I^{a} \otimes Y_{j})\nonumber, \\
t_{ij} &=\textmd{tr}\rho^{ab}(X_{i}\otimes Y_{j}).
\end{eqnarray}
By making use of the above representation of bipartite quantum states, the measure of non-classical correlations is given by
\begin{equation}
\mathcal{Q}(\rho^{ab})=\frac{1}{4}  \sum_{i=m}^{m^2-1}|\Lambda_{i}|,
\end{equation}
where $\Lambda_{i}$ are the eigenvalues of the criterion matrix $\Lambda = T T^{t} - \vec{y}^{t} \vec{y} \vec{x} \vec{x}^{t}$ in decreasing order. The motivation behind the definition of this measure and details of its derivation can be found in Ref. [6].

Negativity is a measure of entanglement that can be straightforwardly calculated for an arbitrary bipartite system in all dimensions. Although we cannot conclude whether a positive partial transpose state (zero negativity state) is entangled or separable in general, negativity is still a reliable measure for all negative partial transpose states [7,8]. For a given bipartite density matrix $\rho^{ab}$, it can be defined as the absolute sum of the negative eigenvalues of partial transpose of $\rho^{ab}$ with respect to the smaller dimensional system,
\begin{eqnarray}
N(\rho^{ab})=\frac{1}{2}\sum_{i}|\eta_{i}|-\eta_{i},
\end{eqnarray}
where $\eta_{i}$ are all of the eigenvalues of the partially transposed density matrix $(\rho^{ab})^{t_{A}}$.

\section{Spin-1 Bose-Hubbard Model}

The Hamiltonian describing the system of spin-1 atoms in an optical lattice is given by [9,10]
\begin{eqnarray}
H &= -t\sum_{\langle ij\rangle, \sigma}(a^{\dag}_{i\sigma}a_{j\sigma}+a_{i\sigma}a^{\dag}_{j\sigma}) +\frac{U_0}{2}\sum_i\hat{n}_i(\hat{n}_i-1)
\nonumber \\
  &+\frac{U_2}{2}\sum_i((\textbf{S}^i_{tot})^2-2\hat{n}_i),
\end{eqnarray}
where $a^{\dag}_{i\sigma}$ $(a_{i\sigma})$ is the creation (annihilation) operator for an atom on site $i$ with $z$ component of its spin being equal to $\sigma=-1, 0, 1$. Here $\hat{n}_i=\sum_{\sigma}a^{\dag}_{i\sigma}a_{i\sigma}$ is the total number of atoms on site $i$ and $\textbf{S}^i_{tot}$ gives the total spin on $i$th lattice site. The parameter $t$ represents the tunneling amplitude, $U_0$ is the on-site repulsion and $U_2$ differentiates the scattering channels between atoms with different $\textbf{S}_{tot}$ values.

From this point on, we assume that the temperature is low enough and the tunneling amplitude $t$ is small so that the overlap between the wavefunctions of the particles in neighboring sites is almost zero. Under these assumptions, the spin-1 Bose-Hubbard Hamiltonian can be treated perturbatively. Second order perturbation theory in $t$ gives the effective Hamiltonian as [10]
\begin{equation}
\frac{H^e_t}{t}=\omega J_z+rI+\tau\sum_{\langle ij\rangle}(\textbf{S}_i\cdot \textbf{S}_j)+\gamma\sum_{\langle ij\rangle}(\textbf{S}_i\cdot \textbf{S}_j)^2.
\end{equation}
In addition to the original spin-1 Bose-Hubbard Hamiltonian, an external magnetic field $\omega$ has been added to the effective Hamiltonian. $\textbf{S}_i$ is the spin operator of the particle on site $i$ with $\textbf{J}=\sum_i\textbf{S}_i$ and $I$ represents the identity operator. In terms of the original Bose-Hubbard Hamiltonian parameters $t$, $U_0$, $U_2$, the effective coupling constants $r$, $\tau$, $\gamma$ for single particle per site are given by
\begin{eqnarray}
r &=\frac{4t}{3(U_0+U_2)}-\frac{4t}{3(U_0-2U_2)}, \nonumber \\
\tau &=\frac{2t}{U_0+U_2}, \nonumber \\
\gamma &=\frac{2t}{3(U_0+U_2)}+\frac{4t^2}{3(U_0-2U_2)}.
\end{eqnarray}
with $r=\tau-\gamma$. In what follows, we will consider the two and three particle cases with a single particle per site.

\subsection{Two particles}

In this setting, the explicit form of the effective Hamiltonian given by Eq. (6) reads
\begin{equation}
H_2=\omega J_z+rI+\tau\textbf{S}_1\cdot \textbf{S}_2+\gamma(\textbf{S}_1\cdot \textbf{S}_2)^2.
\end{equation}
Using the identity $\textbf{S}_1\cdot \textbf{S}_2=(J^2-S^2_1-S^2_2)/2$, the two particle Hamiltonian $H_2$ can be written in the total spin basis as
\begin{equation}
H_2=\omega J_z+\frac{\tau}{2}(J^2-4I)+\frac{\gamma}{4}(J^2-4I)^2+rI,
\end{equation}
where the energy eigenvalues are determined as $E_{JM}=\omega M+\tau(j(j+1)-2)/2+\gamma[(j(j+1)-4)^2-4]/4$.
The density matrix of our system at finite temperature $T$ can be written as
\begin{equation}
\rho_T=\frac{e^{-\beta H}}{Z},
\end{equation}
with the partition function of the system is given by $Z=tr(e^{-\beta H})=e^{-\beta\tau}[2\cosh\beta\tau(1+2\cosh\beta\omega)+\e^{-\beta(3\gamma-2\tau)}+2\e^{-\beta\tau}\cosh2\beta\tau]$ and $\beta=1/T$ with Boltzmann constant $k_B=1$.

In Fig. 1 (a) and (b), we present our results related to the thermal entanglement and quantum correlations in the system of two particles as a function of $\tau$ when $\gamma=\omega=1$ for $T=0.05, 0.5, 1$. Leggio et al. have recently discussed the behavior of thermal entanglement in this model, revealing a connection between the different phases of entanglement and the energy level crossings in the ground state of the system [11]. We demonstrate here that not only the negativity but also the non-classical correlations of the system experience two sharp transitions at points $\tau=0.5$ and $\tau=4$ when the temperature is sufficiently low. Examining the Fig. 1 (c), it is not difficult to see that these sharp transitions are connected with the appearance of energy level crossings in the ground state of the system. In fact, ground state crossings occur at the points $2\tau=\omega$ and $\tau=\omega+3\gamma$, and the connection between the crossings and the considered correlation measures is independent of the values of $\gamma$ and $\omega$. We also note that when $\tau<0$, non-classical correlations in the system grows and reaches to a constant value in this regime as the temperature is increased.

\subsection{Three particles}

When it comes three particles, the effective Hamiltonian with periodic boundary conditions takes the form
\begin{eqnarray}
H_3&=\omega J_z+rI+\tau(\textbf{S}_1\cdot\textbf{S}_2+\textbf{S}_2\cdot\textbf{S}_3+\textbf{S}_3\cdot\textbf{S}_1) \nonumber \\
&+\gamma[(\textbf{S}_1\cdot \textbf{S}_2)^2+(\textbf{S}_2\cdot \textbf{S}_3)^2+(\textbf{S}_3\cdot \textbf{S}_1)^2].
\end{eqnarray}
Similarly to the case of two particles, we straightforwardly obtain the energy eigenvalues of the Hamiltonian and the thermal density matrix $\rho_{T}$ to evaluate the negativity and non-classical correlations in the system. In this case, negativity and quantum correlations are calculated considering the bipartition of $3\otimes9$, that is, we look at the correlations between the first particle and the remaining two particles in the system. Despite the fact that we do not investigate the multipartite non-classical correlations, one can indeed use tripartite negativity defined in Ref. [12] to analyze the multipartite entanglement. It is easy to see that, due to the symmetry of the considered system, the tripartite negativity reduces to usual negativity which is calculated by taking the partial transpose with respect to any of the three qubits.

Fig. 2 (a) and (b) display our results on the thermal entanglement and quantum correlations in the system of three particles with periodic boundary conditions as a function of $\tau$ when $\gamma=\omega=1$ for $T=0.05, 0.5, 1$. The low lying energy levels and their crossing points are also shown in Fig. 3 (c). Looking at the figures, we observe that the two sudden jumps of negativity and quantum correlations correspond to the crossings of the energy levels in the ground state of the system at $\tau=1/3$ and $\tau=2/3$. We note that, different from the case of two particles, negativity and quantum correlations do not show a decreasing behavior about the second transition point, $\tau=2/3$, in case of three particles. Moreover, the plateau occurring after the first transition here is considerably shorter as compared to the two particle case. Lastly, the revival of non-classical correlations with increasing temperature can also be seen when $\tau<0$.

\section{Conclusion}
In summary, we have investigated the thermal quantum correlations and entanglement in a spin-1 Bose-Hubbard model with two and three particles with periodic boundary conditions. Our results demonstrate that both the behavior of thermal quantum correlations and entanglement spotlight the energy level crossings in the ground state of the system. Despite the fact that our discussion is limited to few particle systems, the connection between the behavior of correlations measures and ground state crossings might have consequences even for real quantum critical systems having large number of particles. Finally, it would be interesting to analyze the relation of some thermodynamical quantities (such as the specific heat) and correlations measures since various non-trivial behaviors of certain thermodynamical quantities might give information about the correlations in the system [13,14].

\section*{Acknowledgements}
This work has been partially supported by the Scientific and Technological Research Council of Turkey (TUBITAK) under Grant 111T232.

\begin{figure}[H]
\begin{center}
\includegraphics[scale=1.25]{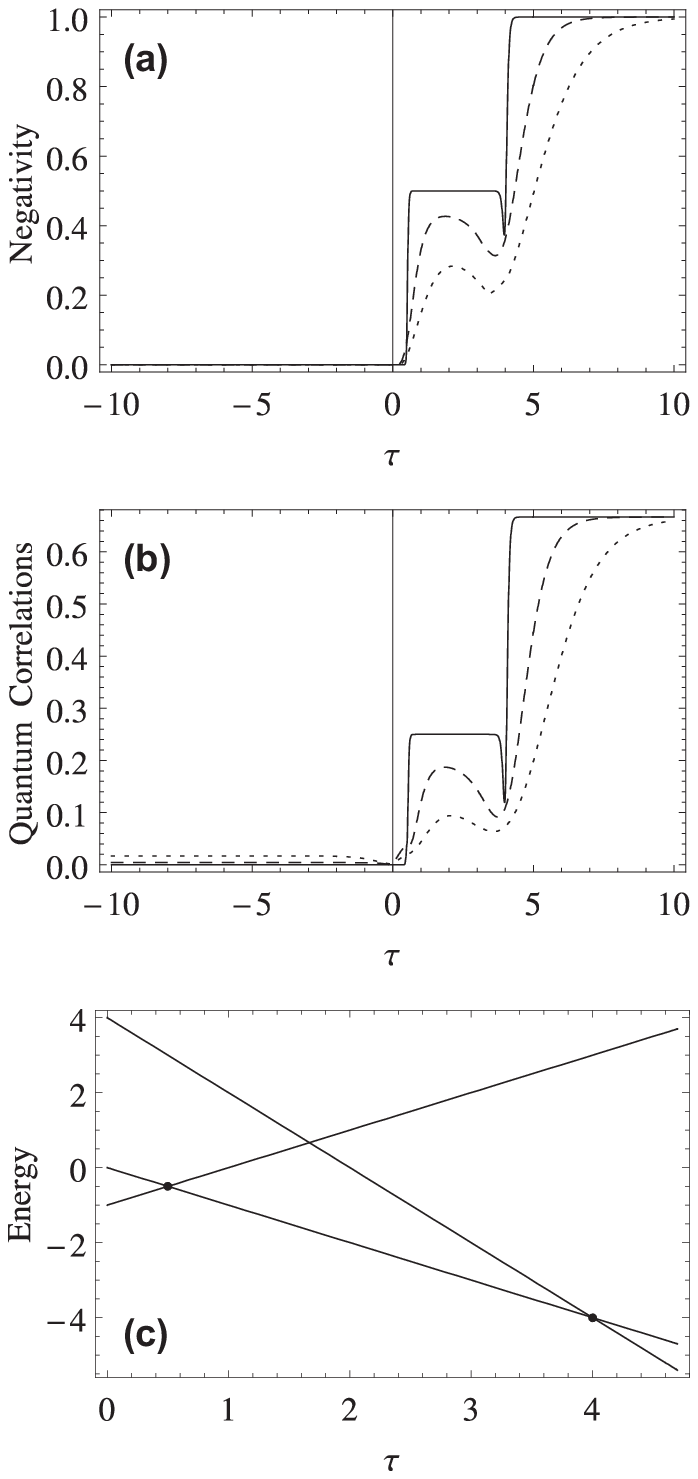}
\caption{The thermal entanglement (a) and quantum correlations (b) of Spin-1 Bose-Hubbard model with two particles as a function of the parameter $\tau$ when $\gamma=\omega=1$ for $T=1$ (dotted line), $T=0.5$ (dashed line) and $T=0.05$ (solid line). The low lying energy levels and their crossings in the ground state of the system are displayed in (c).}
\end{center}
\end{figure}

\begin{figure}[H]
\begin{center}
\includegraphics[scale=1.25]{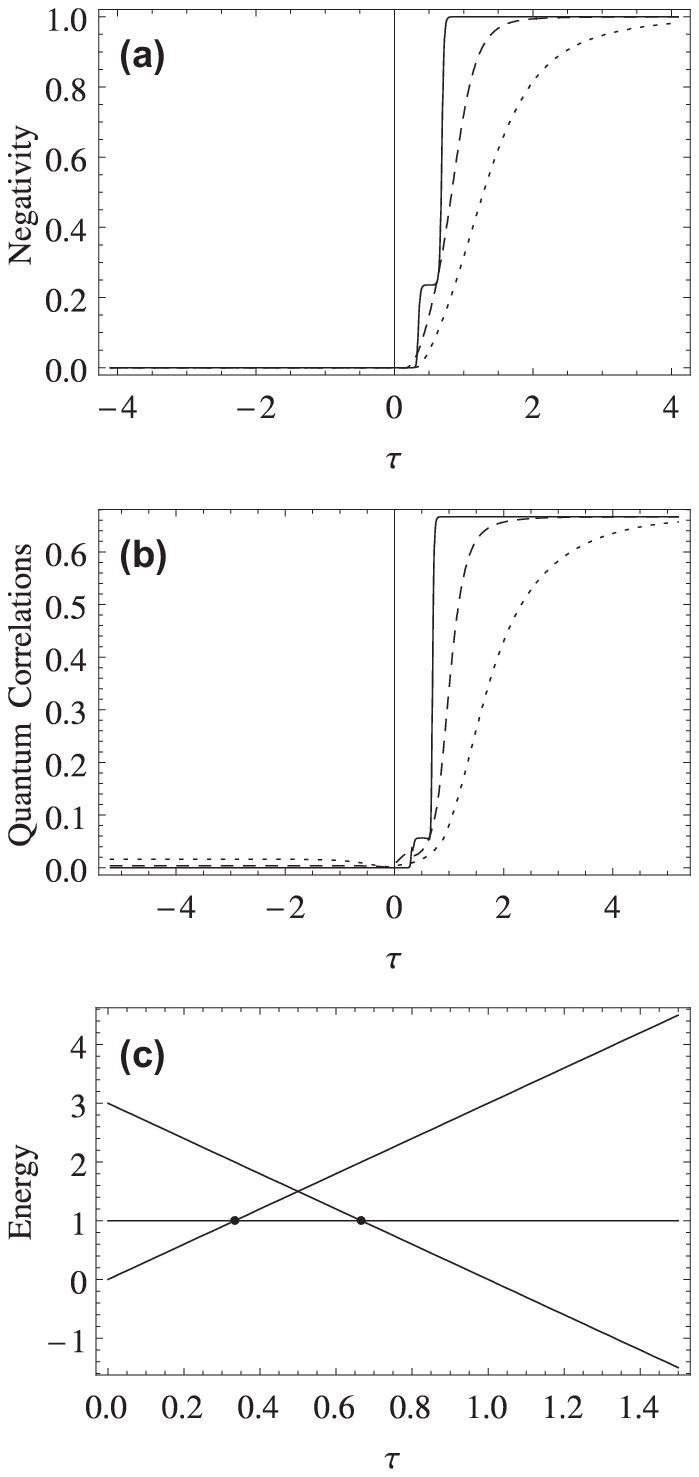}
\caption{The thermal entanglement (a) and quantum correlations (b) of Spin-1 Bose-Hubbard model with three particles as a function of the parameter $\tau$ when $\gamma=\omega=1$ for $T=1$ (dotted line), $T=0.5$ (dashed line) and $T=0.05$ (solid line).The low lying energy levels and their crossings in the ground state of the system are displayed in (c).}
\end{center}
\end{figure}

\end{document}